\documentclass[conference]{IEEEtran}
\IEEEoverridecommandlockouts
\usepackage{cite}
\usepackage{amsmath,amssymb,amsfonts}
\usepackage{algorithmic}
\usepackage{graphicx}
\usepackage{subcaption}
\usepackage{textcomp}
\usepackage{xcolor}
\usepackage{url}
\usepackage{float}

\newtheorem{remark}{Remark}
\newtheorem{definition}{Definition}

\newtheorem{prop}{Proposition}

\def\BibTeX{{\rm B\kern-.05em{\sc i\kern-.025em b}\kern-.08em
    T\kern-.1667em\lower.7ex\hbox{E}\kern-.125emX}}
\begin{document}

\title{Achieving Low Latency at Low Outage: 
\\ Multilevel Coding for mmWave Channels
\thanks{The research at UCLA was supported in part by the U.S. National Science Foundation (NSF) under grant no. \mbox{ECCS-2229560}. The work of M. Cardone was supported in part by the NSF under grant no. \mbox{CCF-2045237}. This material is also based upon work supported by the NSF under grant no. \mbox{CNS-2146838} and is supported in part by funds from federal agency and industry partners as specified in the Resilient \& Intelligent NextG Systems (RINGS) program.}
}

%\author{\IEEEauthorblockN{Mine Gokce Dogan}
%\IEEEauthorblockA{\textit{University of California, Los Angeles} \\
%Los Angeles, CA 90095, USA \\
%minedogan96@g.ucla.edu}
%\and
%\IEEEauthorblockN{Martina Cardone}
%\IEEEauthorblockA{\textit{University of Minnesota}\\
%Minneapolis, MN 55455, USA \\
%mcardone@umn.edu}
%\and
%\IEEEauthorblockN{Christina Fragouli}
%\IEEEauthorblockA{\textit{University of California, Los Angeles}\\
%Los Angeles, CA 90095, USA \\
%christina.fragouli@ucla.edu}
%

 \author{%
   \IEEEauthorblockN{Mine Gokce Dogan\IEEEauthorrefmark{1},
                     Jaimin Shah\IEEEauthorrefmark{2},
%                     Stefan M.~Moser\IEEEauthorrefmark{1}\IEEEauthorrefmark{4},
                     Martina Cardone\IEEEauthorrefmark{2},
                     Christina Fragouli\IEEEauthorrefmark{1},
                     \\ Wei Mao\IEEEauthorrefmark{3},
                     Hosein Nikopour\IEEEauthorrefmark{3},
                     and Rath Vannithamby\IEEEauthorrefmark{3}
                     }
   \IEEEauthorblockA{\IEEEauthorrefmark{1}%
                     University of California, Los Angeles
                     Los Angeles, CA 90095, USA
                     minedogan96@g.ucla.edu, christina.fragouli@ucla.edu}
    \IEEEauthorblockA{\IEEEauthorrefmark{2}%
                     University of Minnesota,
                     Minneapolis, MN 55455,
                     \{shah0732, mcardone\}@umn.edu}
  \IEEEauthorblockA{\IEEEauthorrefmark{3}%
                    Intel Corporation,
                    {Santa Clara, CA 95054}
                    \{wei.mao, hosein.nikopour, rath.vannithamby\}@intel.com}
 }

%\and
%\and
%\and
%\and
%\and

\maketitle

\begin{abstract}
Millimeter-wave (mmWave) spectrum is expected to support data-intensive applications that require ultra-reliable low-latency communications (URLLC).
However, mmWave links are highly sensitive to blockage, which may lead to disruptions in the communication. Traditional techniques that build resilience against such blockages (among which are interleaving and feedback mechanisms) incur delays that are too large to effectively support URLLC. This calls for novel techniques that ensure resilient URLLC. In this paper, we propose to deploy multilevel codes over space and over time. These codes offer several benefits, such as they allow to control what information is received and they provide different reliability guarantees for different information streams based on their priority. We also show that deploying these codes leads to attractive trade-offs between rate, delay, and outage probability.
A practically-relevant aspect of the proposed technique is that it offers resilience while incurring a low operational complexity.
\end{abstract}

\section{Introduction}
Next generation wireless networks are expected to support a wide range of data-intensive applications that require {\em \mbox{ultra-reliable} \mbox{low-latency} communications} (URLLC). Examples include cloud gaming and live stream $360^\circ$ virtual reality. These applications impose strict Quality of Service (QoS) requirements: packet delay budgets of $50$~ms, packet error rates of $10^{-3}$, and data rates up to $80$~Mbps~\cite{3gpp.26.925}. URLLC are also required for \mbox{mission-critical} applications, such as autonomous driving, factory automation, and remote surgery.

A key enabling technology that can support the URLLC use cases leverages the \mbox{{\em millimeter-wave}} (mmWave) spectrum.
Despite this promising aspect of mmWave communications, it is well known that mmWave links are highly sensitive to blockage and communication can get disrupted.
Traditional techniques that offer resilience against blockages use interleaving and feedback mechanisms.
However, these come at the cost of low information rate, increased latency, or low~reliability. %Hence, they are not suitable to support URLLC.

In this paper, we propose to deploy {\em multilevel codes}~\cite{Roche97,Yeung99} for resilient URLLC. In particular, we encode the source sequences in packets and send them over multiple network paths (that may exist between a source and a destination) and over multiple time slots. We do this by assuming the knowledge of the link blockage probabilities. These probabilities can be estimated through accurate models in advance~\cite{Jain,Gapeyenko17,Wang17,Raghavan19}.
Our proposed transmission schemes have the following advantages.

{\em First,} they are {\em proactive}, i.e., they build resilience in advance without an a priori knowledge of the blockages.
This ensures communication guarantees with no additional delay, even if blockages diminish network resources. Thus, proactive mechanisms are suitable for \mbox{delay-sensitive} applications which may require latency as low as a few milliseconds~\cite{3gpp.26.925}.

{\em Second,} multilevel codes allow to {\em control what information is received} even if only a subset of the paths are available to operate. This is a challenging task because once a blocker interrupts a communication link, it causes that link to become unavailable for a certain duration. 
Thus, only a subset of the paths might be available while operating the network, and we do not know in advance which ones. We cannot simply ``average out" these events while providing reliability and latency guarantees for \mbox{delay-sensitive} applications. For example, consider a network with $6$ paths, all with blockage probability $0.3$. Assume that once a blocker interrupts a path, it continues to interrupt that path for $500$ ms. When this network starts to operate, any $2$ paths can get blocked with probability $0.32$. That means, only $4$ of the paths (and we do not know which ones) can be operational for $500$ ms. If we simply send uncoded data, we cannot control the received information when some paths are operational for a certain duration\footnote{If we send $6$ independent information streams, one through each path, we will have no control on which information stream will be received.}. 

{\em Third,} multilevel codes offer different reliability guarantees to different information streams based on their {\em priority}. This is particularly important for data-intensive applications, in which more relevant information streams need to be received with a higher probability and/or a lower latency, and/or higher rate.

{\em Fourth,} multilevel codes do not have a single threshold of failure. They provide a graceful performance degradation: if less than the expected amount of blockages occur, we can leverage this to increase the information rate; and if more blockages occur, the information rate will gradually decrease.

The aforementioned advantages come with a certain challenge: the operational complexity of multilevel codes increases with the number of paths utilized and the code duration.

\smallskip

\noindent \textbf{Related Work.}
Several works in the literature offer resilience against link outages by taking reactive approaches~\cite{Jeong,Barati,MineMilcom}. However, such reactive mechanisms add the feedback latency and the complexity of identification and adaptation. Several works proposed proactive approaches for resilience~\cite{Va,Mine22}, but they are different from our work as we propose coding schemes to control what information is received, and to accommodate different reliability requirements of different information streams. In~\cite{MineGlobecom}, the authors proposed \mbox{low-complexity} proactive mechanisms for mmWave networks by deploying multilevel codes over space (i.e., across multiple paths). The authors then extended this work to scenarios in which the path blockage probabilities are unequal~\cite{MineMILCOM23}. These works are different from our work as: (i) the operation of their schemes relies on a multipath environment, which is not always practical; and (ii) they focus on the \mbox{rate-outage probability} \mbox{trade-off} without any delay requirements. On the contrary, in this paper: (i) we propose to deploy multilevel codes \textit{both} over space and time, which allows to deploy them in networks that do not support a multipath environment; (ii) we consider time correlation of blockages; and (iii) we consider the \mbox{trade-off} between the rate, delay, and outage probability. The extended version of the Related Work is delegated to~Appendix~\ref{appendix:related_work}.

\noindent \textbf{Contributions and Paper Organization.}
In Section~\ref{sec:System Model and Background}, we provide an overview on the 1-2-1 model, on the erasure codes, and on the symmetric multilevel codes. In Section~\ref{sec:channel_analysis}, we analyze the channel under the considered blockage model. In particular, we derive the probability mass function (PMF) of the number of received packets, and we further analyze this distribution. 
%of the number of received packets. 
In Section~\ref{sec:proposed_coding_scheme}, we propose proactive transmission mechanisms for mmWave networks. In particular, we propose to deploy symmetric multilevel codes over space and time. Towards achieving an attractive \mbox{trade-off} between the rate and a graceful performance degradation, we propose an optimization formulation to choose our design parameters. We also present a \mbox{low-complexity} coding scheme that aims at approximating well the aforementioned \mbox{trade-off}. In Section~\ref{sec:NumericalEvaluation}, we numerically evaluate the performance of our schemes and compare them with an alternative scheme. In particular, we investigate the \mbox{trade-off} between the rate, delay, and outage probability. Our evaluations show that: (i) the proposed schemes achieve a more attractive \mbox{trade-off} between the rate, delay, and outage probability by providing a more graceful performance degradation compared to the alternative scheme; and (ii) our complexity reduction technique gives a comparable performance, while significantly reducing the code complexity. Finally, in Section~\ref{sec:Concl} we conclude the paper.
%Millimeter-wave (mmWave) networks offer high data rates, making them suitable for applications like Augmented Reality (AR) and Virtual Reality (VR), which demand data rates between 100 Mbps and a few Gbps and low latency between 1 ms and 10 ms.(cite??) However, mmWave systems are prone to blockages, including those caused by the human body, leading to wireless channel alternate between the blocked and non-blocked LoS states. These blockages can be detrimental to ultra-reliable and low-latency communication (URLLC) applications. 3GPP 5G NR standared has adopted design of TTI towards lower latency and more efficient use of radio resources. It incorporates a scalable TTI design, with slot durations ranging from 62.5µs to 1ms. However depending on nature of blocker when block duration is of order of 10ms or more QoS and reliability is impacted greatly.

\section{System Model and Background}
\label{sec:System Model and Background}
\noindent \textbf{Notation.} $[a \!:\! b]$ is the set of integers from $a$ to $b > a$, and $|\cdot|$ is the cardinality for sets; $*$ denotes the convolution operation. For a vector $v$, we denote with $\|v\|$ the $\ell_2$-norm of~$v$. 
%The convolution operation is denoted by $*$.

We build on the \hbox{1-2-1} network model, which was introduced to study the \mbox{information-theoretic} capacity of mmWave networks~\cite{ezzeldin}. The model abstracts away the physical layer component and focuses on modeling  the {\em directivity} characteristic of mmWave communications: mmWave nodes perform beamforming with narrow beams to compensate the path loss. 

We consider a \hbox{1-2-1} network with $N$ relays that assist the communication between the source \mbox{(node $0$)} and the destination (node $N\!+\!1$). The relays can operate either in \hbox{full-duplex} or \hbox{half-duplex} mode. Two nodes steer their beams towards each other to activate a link that connects them (called a \hbox{1-2-1} link~\cite{ezzeldin}). At any given time, the source and the destination can steer their beams towards $H$ nodes ($H$ denotes the number of \mbox{edge-disjoint} paths in the network), whereas the relays can transmit to (and receive from) at most another node\footnote{Our results hold even if relays have multiple transmit and receive beams.}.
%The authors in~\cite{ezzeldin} approximated the unicast capacity of an \mbox{$N$-relay} Gaussian \hbox{1-2-1} network to within an additive gap that only depends on $N$. Moreover, they showed that an optimal beam schedule that achieves the approximate capacity can be computed in polynomial time in $N$ by a linear program, both for \hbox{full-duplex} and \hbox{half-duplex} modes. In~\cite{Mine22}, authors extended these results to scenarios when the network links experience blockage. 
%In the scope of this paper, We focus on a network model that has $H$ number of \mbox{edge-disjoint} paths between source and destination node all having equal capacity i.e., in the case of unit capacity channel, source node can simultaneously transmit $H$ packets(bits) on $H$ \mbox{edge-disjoint} path and  the destination node can receive $H$ packet(bit) simultaneously to decode the information. 

\noindent \textbf{Link Blockage Probabilities.} We build on the \textit{existence} of accurate models that estimate the link blockage 
%(failure) 
probabilities in mmWave networks~\cite{Jain,Gapeyenko17,Wang17,Raghavan19}. 
%In particular, 
These works model the blocker arrival process as a Poisson point process (PPP). In particular, the intensity $\alpha_{j,i}$ of the Poisson process for the link from node \hbox{$i \in [0\!:\!N]$} to node \hbox{$j \in [1\!:\!N{+}1]$} is $\alpha_{j,i} = \lambda_{j,i}d_{j,i}$, where: (i) $\lambda_{j,i}$ is proportional to the blocker density and velocity, and to the heights of the blockers, the receiver and the transmitter~\cite{Jain}; and (ii) $d_{j,i}$ is the distance between nodes $i$ and $j$. 

\begin{figure}
	\centering
    \includegraphics[width=0.7\columnwidth]{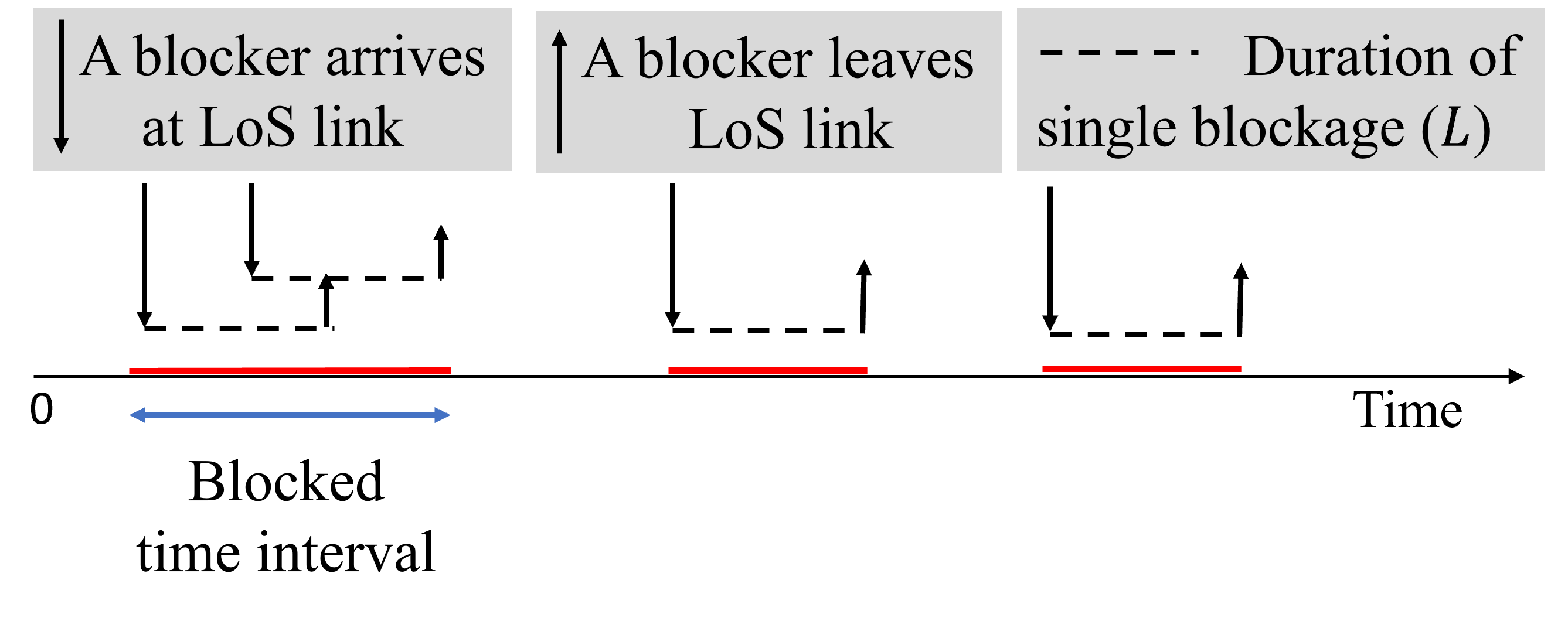}
    \vspace{-0.1in}
     \caption{Blockage model illustration over a single LoS link.}
     \label{fig:link bloackage}
     \vspace{-0.2in}
\end{figure}

Similarly, in this paper we assume a PPP for the blocker arrivals. If a blocker interrupts a \hbox{line-of-sight} (LoS) link, it continues to interrupt that link for the consecutive $L$ time slots, where $L$ is a constant value. 
%\textcolor{blue}{MD: is it okay if I emphasize that $L$ is a constant value instead of a random value?} 
In this work, we assume uncorrelated blockages across different links.
%\textcolor{blue}{MD: One of the reviewers wasn't sure if a single blocker is considered for a two-hop cascade TX-RELAY-RX. Should I state that we consider a multi-hop setting and links between relays can also get blocked?} 
In Fig.~\ref{fig:link bloackage}, we illustrate our blockage model for a single link. We allow for overlaps of blockages as shown in Fig.~\ref{fig:link bloackage}. That is, if a blocker interrupts a link, in the meantime another blocker can start to interrupt the same link. This increases the total blockage duration as shown with a red block in Fig.~\ref{fig:link bloackage}.

%In \mbox{mm-Wave} networks, links are susceptible to blockage and the link blockage probabilities can be estimated in advance through existing models(cite??). (cite??) has devloped model to capture the effect of dynamic blockage as well as the temporal consistency of the channel at mmWave frequencies showing that the blockage effects produce an alternating renewal process with exponentially distributed non-blocked intervals, and blocked durations that follow the general distribution. However in this work we modeled bloacker with poison renewel process having constant bloackage duration where the intensity of blockers entering the LoS blockage zone on path $h$ be  given as,(cite??)
%\begin{align*}
%    \lambda_h = \lambda_{S,h} r_h d_h    
%\end{align*}
%where (i) $\lambda_{S,h}$ is the constant arrival
%intensity of blockers; (ii) $r_h$ is the effective length of LoS blockage zone; and (iii) $d_h$ is the diameter of blockers on path $h$. 

%Thus, the inter-arrival time between blockers has exponential distribution with average rate of $\lambda_h$, and once the bloackage happened on path $h$ it will remain for a constant amount of time denoted as $L_h$

\noindent \textbf{Erasure Codes.} An erasure code is a forward error correction code that assumes packet erasures (losses)~\cite{Reed60,huffman_pless_2003}. An erasure code $(n, k)$ transforms $k$ information packets into $n$ encoded packets such that the original message is reconstructed if any $k$ packets (out of the $n$ transmitted packets) are received. This results in an information rate of $k/n$. An erasure code supports a given number of blockages: we experience ``outage'' if the number of blockages is higher than the design (less than $k$ packets are received, resulting in a zero information rate); and we succeed if there are fewer blockages than the design (at least $k$ packets are received, resulting in an information rate of $k/n$). Thus, erasure codes do not offer a graceful performance degradation. Moreover, even if we succeed, experiencing fewer blockages does not increase the information rate. %We next formally define the average rate and the outage probability of erasure codes.

\noindent \textbf{Multilevel Diversity Coding (MDC).} MDC is a classical coding scheme that provides a graceful performance degradation. It encodes i.i.d. source sequences to accommodate different reliability requirements of different source sequences. MDC can be designed in two ways: symmetric and asymmetric. In this paper, we build our schemes on symmetric MDC.

In symmetric MDC~\cite{Roche97,Yeung99}, $H$ i.i.d. source sequences are considered. They have certain levels of importance, ordered from $1$ (the most important) to $H$ (the least important). They are encoded into $H$ descriptions using $H$ encoders. These descriptions are sent to $H$ decoders, each through a different channel. There are $H$ levels and the decoders are assigned with ordered levels. Each decoder has access to a subset of the descriptions, and its level depends on the number of descriptions to which it has access. The encoders produce the descriptions such that a decoder at level $h$ (i.e., has $h$ available descriptions) can reconstruct the $h$ most important source sequences, $h \in [1\!:\!H]$. For symmetric MDC, superposition coding is optimal~\cite{Roche97,Yeung99}. That is, each source sequence is encoded separately, and the descriptions are created by concatenating the encoded sequences appropriately. The next example illustrates the \mbox{$3$-level} MDC and its potential benefits. 

\begin{figure}
	\centering
    \includegraphics[width=0.28\columnwidth]{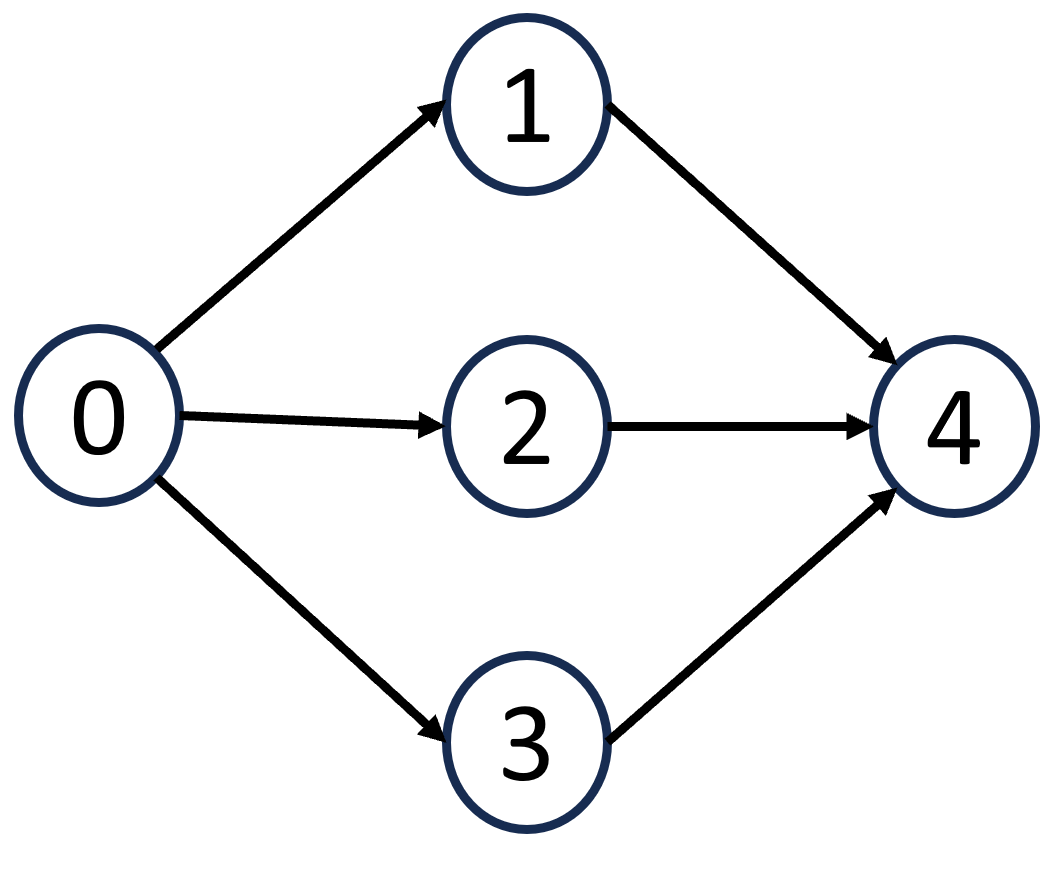}
     \caption{An example network with $N=3$ relays.}
     \label{fig:example_network1}
     \vspace{-0.2in}
\end{figure}

\begin{figure}
	\centering
    \includegraphics[width=.7\columnwidth]{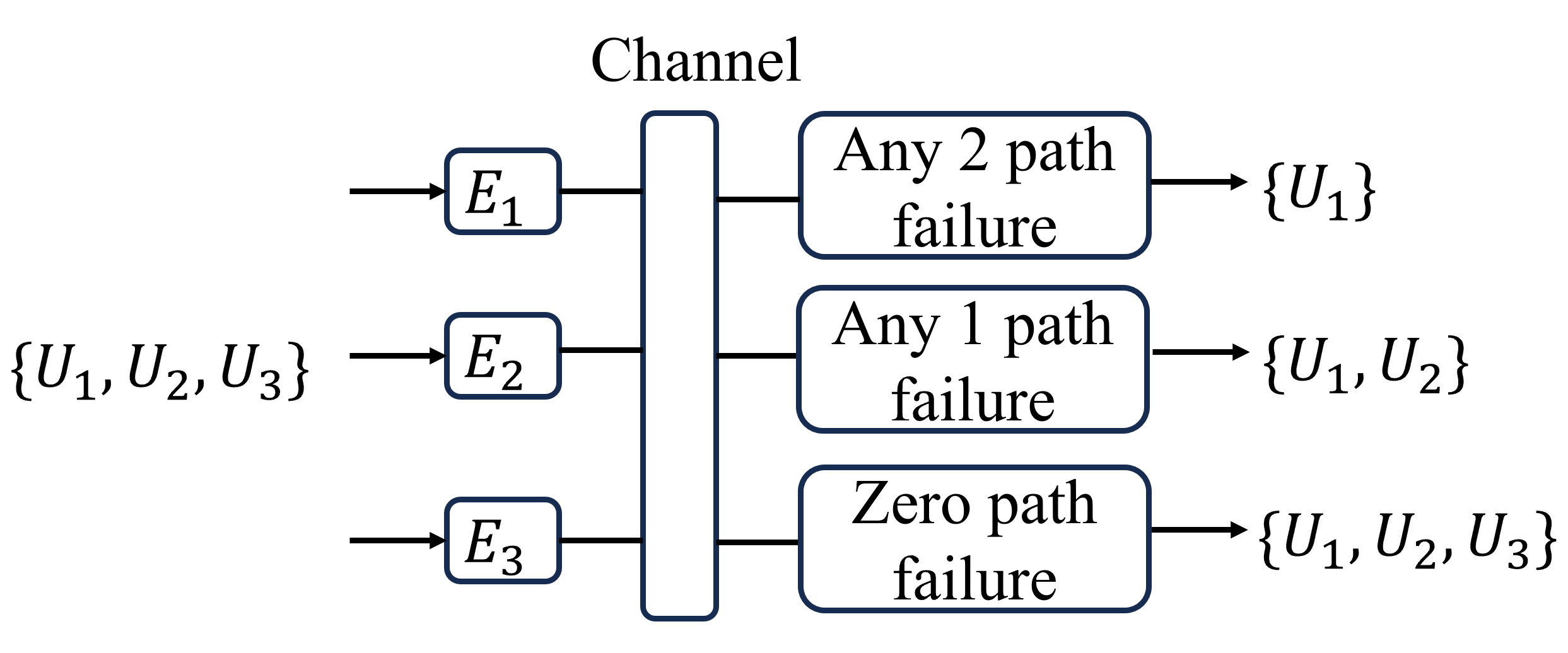}
     \caption{$3$-level symmetric multilevel code setting.}
     \label{fig:symmetric_setting}
     \vspace{-0.2in}
\end{figure}

\noindent \textit{Example 1.} Consider the network in Fig.~\ref{fig:example_network1} that has $H=3$ \mbox{edge-disjoint} paths connecting the source (node $0$) to the destination (node $4$). 
We let $U_i, i \in [1\!:\!3]$ be the i.i.d. source sequences, ordered with decreasing importance. They are encoded by $H=3$ encoders, and each description (denoted by $E_i, i \in [1\!:\!3]$) is sent through a different path. In Fig.~\ref{fig:symmetric_setting}, we show the setting of a $3$-level symmetric multilevel code over this network.
The goal is to reconstruct $U_i, i \in [1\!:\!h],$ if \textit{any} $h$ paths succeed (or equivalently, any $H-h$ paths fail).
\hfill $\square$

\noindent \textbf{Performance Metrics.} We assess the proposed coding schemes performance through the performance metrics below.

\noindent \textbf{1) Outage Probability.} 
As discussed, a single erasure code can support only up to a certain number of packet losses. For a higher number of packet losses, the network experiences outage. The probability of outage is defined as follows.
\begin{definition}\label{def:outage_prob}
The outage probability of an erasure code $(n,k)$ is defined as,
\begin{align}\label{eq:Pout}
P_{\rm{out}} = P(X <  k),
\end{align}
where the random variable $X$ denotes the total number of packets received by the destination. \hfill $\square$
\end{definition}
As we discuss in Section~\ref{sec:proposed_coding_scheme}, multilevel codes can be designed by combining multiple erasure codes. Thus, they do not have a single outage probability: there is a different outage probability for every erasure code combined by the multilevel code.
%\textcolor{blue}{MD: Is this the way that you would like me to define the outage probability of multilevel codes?}

\noindent \textbf{2) Average Rate.} Our second performance metric is the average information rate of an erasure code. 
\begin{definition}\label{def:average_rate}
The \textit{average} information rate of an erasure code $(n,k)$ is defined as,
\begin{align}\label{eq:average_rate_singlecode}
R_{\rm{E,(n,k)}} = \frac{k}{n} \left( 1 - P_{\rm{out}} \right ),
\end{align}
where $P_{\rm{out}}$ is the outage probability in Definition~\ref{def:outage_prob}. \hfill $\square$
\end{definition}
Since a multilevel code can be designed by combining multiple erasure codes (see Section~\ref{sec:proposed_coding_scheme}), its average rate is equal to a weighted sum of the average rates of the erasure codes that are combined. It is formally presented in Definition~\ref{def:avg_rate_mc}.

\noindent \textbf{3) Delay.}
The final performance metric is the delay, which quantifies the amount of time needed to transmit the source sequences. As we discuss in Section~\ref{sec:proposed_coding_scheme}, we propose to deploy the coding schemes over time: we first encode the source sequences and then transmit the encoded sequences over $T$ time slots, where $T$ denotes the code duration. We assume that each time slot lasts for one transmission time interval (TTI) denoted by $t_d$ (e.g., $t_d=250~\mu$s~\cite{Rinaldi21}). Thus, the delay of every coding scheme considered in this paper is equal to $Tt_d$. We choose the value of $T$ according to the latency constraints.

\section{Channel Analysis}\label{sec:channel_analysis}
In this section, we analyze the channel, and we derive the PMF of the number of received packets.
%In this section, we analyze the channel for the blockage model described in Section~\ref{sec:System Model and Background}. Particularly, in Section~\ref{subsec:blockage_prob} we derive the PMF of the number of received packets under the considered blockage model. Then, in Section~\ref{sec:BlockDur} we investigate the effect of the blockage duration on the number of received (or lost) packets, which in turn affects the \mbox{rate-outage probability} \mbox{trade-off}.

%\subsection{PMF of the Number of Received Packets}\label{subsec:blockage_prob}
We consider a \mbox{1-2-1} network with $H$ \mbox{edge-disjoint} paths. As we discuss in Section~\ref{sec:proposed_coding_scheme}, we encode the source sequences in packets and transmit the packets over $T$ time slots. Each time slot $t_k$, \hbox{$k \in [1\!:\!T]$} lasts for one TTI. The blocker arrival process on path $p_j$ is a PPP with intensity $\alpha_j$ per TTI for $j \in [1\!:\!H]$. Thus, the number of blockers that interrupt path $p_j$ at time slot $t_k$, $k \in [1\!:\!T]$ has a Poisson distribution with parameter $\alpha_j$. Since the PPP has independent increments and each time slot is a disjoint interval in time, a new blockage event can independently start on path $p_j$ (i.e., at least one new blocker interrupts the path) at every time slot with probability,
\begin{equation}\label{eq:eps_alpha}
\varepsilon_j = 1-{\rm{e}}^{-\alpha_j},~j \in [1:H].
\end{equation}
%{\color{magenta} JS: Or We can simply define $\alpha_j$ as the number of blockers per TTI}.\textcolor{blue}{MD: Do you mean as the density?} 
As discussed in Section~\ref{sec:System Model and Background}, once a blocker interrupts a path, it continues to interrupt the path for $L$ time slots. In the rest of this paper, we assume that blockage events can occur only at the beginning of a time slot. This implies that  an entire packet is either received or lost; we cannot receive a partial packet.
% Throughout the remainder of this paper, we assume that if a blockage event starts on a path at a particular time slot (i.e., if at least one new blocker interrupts the path), it starts in the beginning of that time slot. If a time slot is blocked, we assume that the packet transmitted during that time slot is lost. Otherwise, it is successfully received.

Let $X$ denote the total number of received packets over $H$ paths through $T$ time slots. We have \hbox{$X = \sum_{j=1}^H X_j$} where $X_j$ denotes the number of received packets on path $p_j$ over $T$ time slots, for $j\in [1:H]$. 
Thus the PMF of $X$, denoted by $P_X$, can be written as,
\begin{equation}
P_X = P_{X_1}*\ldots*P_{X_H},
\end{equation}
where $P_{X_j}$ is the PMF of $X_j$, $j \in [1\!:\!H]$ and it is derived in the following proposition (proof in Appendix~\ref{appendix:proof_pmf_prop}). %\textcolor{blue}{MD: One of the reviewers thought that the transmission was non-continuous, should I emphasize that the transmission is continous but the code duration is $T$ time slots thus we encode a different set of source sequences at every $T$ time slots.}
\begin{prop}\label{prop:pmf}
Consider a 1-2-1 network with $H$ \mbox{edge-disjoint} paths. Let $T$ denote the code duration and $L$ denote the blockage duration in time slots, such that $L \geq T$.  Then, $P_{X_j}$, $j\in[1:H]$ is given by:

\noindent $\bullet$ 
%For $r = 0$, 
The probability \hbox{$P_{X_j}(0) = P(X_j=0)$} is
\begin{equation*}\label{pmf_each_path2}
P_{X_j}(0)=\varepsilon_j+\sum_{i=1}^{T} (1-\varepsilon_j)^{i} \varepsilon_j^{\min\{T-i,1\}}\left(1-(1-\varepsilon_j)^{L-i}\right),
\end{equation*}
where $\varepsilon_j$ is defined in~\eqref{eq:eps_alpha}. 
%{\color{orange}WM: Can we use a different variable here? $k$ was used for the code parameters in \eqref{eq:average_rate_singlecode} and it is easy to cause confusion.}\textcolor{blue}{MD: I changed $k$ to $r$, please let me know if it looks good.}

\noindent $\bullet$ For \hbox{$0 < r \leq T$}, the probability $ P_{X_j}(r) = P(X_j=r)$ is
\begin{equation*}\label{pmf_each_path1}
P_{X_j}(r) \!=\! \sum_{i=0}^{T-r} (1\!-\!\varepsilon_j)^{r+i} \varepsilon_j^{\min\{T-r-i,1\}} (1\!-\!\varepsilon_j)^{L-1-i} \varepsilon_j^{\min\{i,1\}}.
\end{equation*}
\end{prop}
\begin{remark}\label{remark:L_T_assumption}
It is reasonable to assume $L \!\geq\! T$ in practice. First, the value of $T$ is constrained by the latency requirements of \mbox{delay-sensitive} applications
%; which may require 
(at most $100$ ms latency~\cite{3gpp.26.925}). The delay of every coding scheme considered in this paper is $Tt_d$ as discussed in Section~\ref{sec:System Model and Background}, thus we constrain the value of $T$ (e.g., $T \leq 400$ for $t_d = 250~\mu$s). 
%On the other hand, 
Second, measurement studies show that the blockage duration is of the order of $100$ ms~\cite{Jain,Gapeyenko17,Wang17,Raghavan19}. In this work, the blockage duration due to a single blocker is $Lt_d$\footnote{Overlapping blockage events can extend the total blockage duration but the blocked intervals are likely to feature a single blocker occluding the path~\cite{Gapeyenko17}.}. For example, this requires $L \geq 400$ for $t_d=250~\mu$s.
\end{remark}

We next show a property of the number of received packets (see Appendix~\ref{app:ChannelPol} for the detailed proof).
\begin{prop}\label{prop:asymptotic}
If (i) \hbox{$\varepsilon_j(T-1) \ll 1$}, or (ii) \hbox{$(1-\varepsilon_j)^L \ll 1$} and \hbox{$\varepsilon_j(T-1)(1-\varepsilon_j)^L \ll 1$}, the following approximation holds for \hbox{$j \in [1\!:\!H]$},
%The following approximation holds for \hbox{$j \in [1\!:\!H]$}, if one of the conditions is satisfied: (i) \hbox{$\varepsilon_j(T-1) \ll 1$}, or (ii) \hbox{$\varepsilon_j(T-1)(1-\varepsilon_j)^L \ll 1$} and \hbox{$(1-\varepsilon_j)^L \ll 1$}.
\begin{equation}\label{eq:prob_approx}
P_{X_j}(0)+P_{X_j}(T) \approx 1.
\end{equation}
Moreover, it always holds that $\lim_{L\to \infty}P_{X_j}(0) = 1$.
\end{prop}

In practice, the conditions in Proposition~\ref{prop:asymptotic} may hold. First, as pointed out in Remark~\ref{remark:L_T_assumption}, latency requirements constrain the value of $T$, and $L$ may take large values as supported by measurement studies. Second, mmWave networks can support short TTI durations, thus it is reasonable to assume that $\varepsilon_j$  does not take large values. If the approximation in~\eqref{eq:prob_approx} holds, the number of received packets on each path is likely to be either $0$ or $T$ at every $T$ time slots. That means, an uncoded transmission performs well for $H\!=\!1$. However, if the approximation in~\eqref{eq:prob_approx} does not hold or if $H> 1$, the number of received packets takes different values. Assume the approximation in~\eqref{eq:prob_approx} holds and $H >1$, then the number of received packets is likely to take values $jT$ for $j\in[0\!:\!H]$ (see Appendix~\ref{app:NumericalEvalChannel} for numerical analysis). In all such cases, MDC provides a graceful performance degradation. 

\section{Proposed Coding Schemes}\label{sec:proposed_coding_scheme}
In this section, we discuss how to deploy multilevel codes. Our coding schemes are largely based on the schemes proposed in~\cite{MineGlobecom}; however, different from those in~\cite{MineGlobecom}, they can also be deployed over time. This allows to reap their benefits also in networks that do not support a multipath environment, and allows to consider correlated blockages over time.
%In this section, we design proactive transmission mechanisms for mmWave networks. In Section~\ref{subsec:mc_desing} we discuss how to deploy symmetric multilevel codes over space and time. Towards achieving an attractive \mbox{trade-off} between the average information rate and a graceful performance degradation, we {\color{magenta}present} an optimization formulation in Section~\ref{subsec:mc_optimization} to choose our design parameters. We then {\color{magenta}present} a \mbox{low-complexity} coding scheme in Section~\ref{subsec:complexity_reduction} to approximate well the aforementioned \mbox{trade-off}. {\color{magenta}The presented coding schemes are largely based on the schemes proposed in~\cite{MineGlobecom}; however, different from those in~\cite{MineGlobecom}, they can also be deployed over time. This characteristic allows to reap their benefits also in networks that do not support a multipath environment.}
%
%\subsection{Symmetric Multilevel Code Design}\label{subsec:mc_desing}

\begin{figure}
	\centering
    \includegraphics[width=0.5\columnwidth]{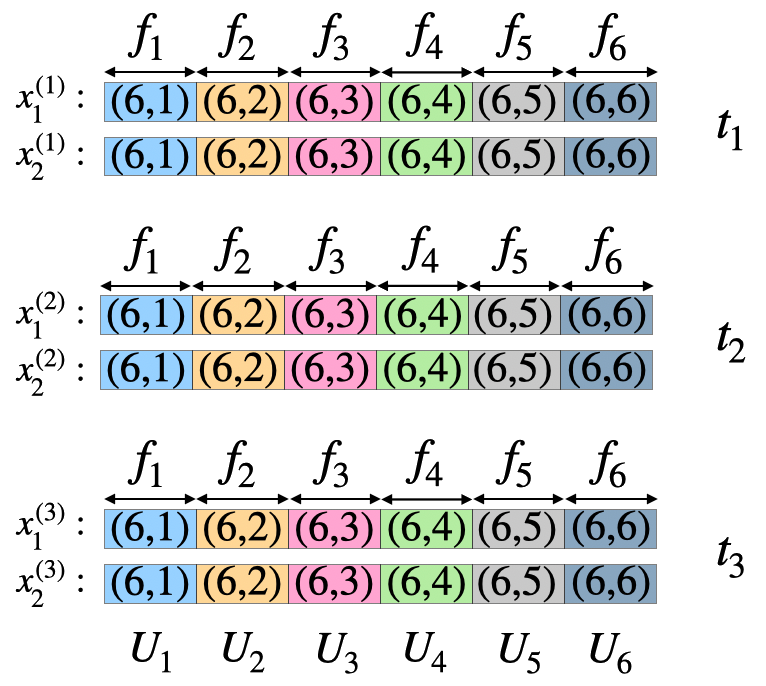}
     \caption{Symmetric multilevel code design for $H\!=\!2$ and $T\!=\!3$.}
     \label{mc_design}
     \vspace{-0.2in}
\end{figure}

We let $H$ denote the number of \mbox{edge-disjoint} paths in the network and $p_{[1:H]}$ is the corresponding set of \mbox{edge-disjoint} paths. We propose to deploy symmetric multilevel codes over paths $p_{[1:H]}$ and over $T$ time slots ($T$ denotes the code duration). Our scheme builds on superposition coding discussed in Section~\ref{sec:System Model and Background}. We consider $HT$ i.i.d. source sequences denoted by $U_1,\ldots,U_{HT}$, which are ordered with decreasing importance. We propose to encode each source sequence $U_i$ with a different rate erasure code $(HT,i)$, $i \in [1\!:\!HT]$. We then concatenate the encoded sequences to create \textit{combined} packets denoted by $x_j^{(k)}$, \hbox{$j \in [1\!:\!H]$}, $k \in [1\!:\!T]$. We transmit $x_j^{(k)}$ through path $p_j \in p_{[1:H]}$ at time slot $t_k$\footnote{As discussed in Section~\ref{sec:System Model and Background}, every combined packet is transmitted during one TTI (denoted by $t_d$) and the transmission duration of $HT$ combined packets is $Tt_d$.}. Every $x_j^{(k)}$ consists of $HT$ components, and each component is created based on a different rate erasure code. We use codes $(HT,i)$, $i \in [1\!:\!HT]$ to create the components. Let $x_{j,i}^{(k)}$, $i \in [1\!:\!HT]$ denote the components of $x_j^{(k)}$. Each $x_{j,i}^{(k)}$ is created based on a code $(HT,i)$. We allocate a packet fraction to each code while creating the combined packets: $f_i$ denotes the fraction of a combined packet allocated to code $(HT,i)$, $i \in [1\!:\!HT]$. In Fig.~\ref{mc_design}, we illustrate our scheme for $H=2$ and $T=3$. In what follows, we refer to the combined packets as packets.

Our scheme guarantees higher reliability to more important source sequences. For example in Fig.~\ref{mc_design}, the most important source sequence $U_1$ is encoded with the most reliable code $(6,1)$ (i.e., has the smallest outage probability). Thus, $U_1$ is successfully decoded if at least one packet is received.
Under this scheme, if $r$ packets are received (out of the $HT$ transmitted packets) for $r \in [1\!:\!HT]$, the information rate is equal to $\sum_{i = 1}^r (i/HT)f_i$. The average information rate $R_{\mathrm{MC}}$ is defined similarly, where MC refers to this proposed scheme.
\begin{definition}\label{def:avg_rate_mc}
The average information rate $R_{\mathrm{MC}}$ of a symmetric multilevel code is,
\begin{equation}\label{eq:avg_rate_mc}
R_{\mathrm{MC}} = \sum_{i\in [1:HT]} \left(f_i P(X \geq i)\frac{i}{HT}\right),
\end{equation}
where $X$ denotes the number of received packets out of the transmitted $HT$ packets.
%; and (ii) $\delta$ denotes the set of indices\footnote{For an erasure code $(n,k)$, its index is equal to $k$, i.e., the number of information packets created by the code.} of the erasure codes that are combined, thus \hbox{$\delta = [1\!:\!HT]$}.
\end{definition}
\begin{remark}
The average information rate $R_{\mathrm{MC}}$ in~\eqref{eq:avg_rate_mc} is equal to a weighted sum of average rates of erasure codes as defined in~\eqref{eq:average_rate_singlecode}. The weights are the packets fractions \hbox{$f_i$, $i \in [1:HT]$}. 
\end{remark}

\subsection{Selection of the Packet Fractions}\label{subsec:mc_optimization}
%An important design question in our scheme is the allocation {\color{magenta}of the packet} fractions $f_i$, $i\in[1:HT]$. 
We propose the following optimization problem, which can be solved with off-the-shelf solvers, to select the packet fractions $f_i$, $i\in[1:HT]$,
%\textcolor{blue}{MD: Should I state that the following optimization problem can be solved with any off-the-shelf solver?}
\begin{align}\label{symmetric_optimization}
\begin{array}{llll}
&  \underset{f}{\max} \ \sum_{i \in [1:HT]} \left(\frac{i}{HT}P(X\geq i)f_i\right)-\mu \lVert f\rVert^2   & \\
& {\text{subject to}} \ \sum_{i \in [1:HT]} f_i=1, \\
& {\text{and}} \ \  \ \ \ \ \ \  f \geq 0,
\end{array}
\end{align}
where: (i) $f$ denotes the vector of the packet fractions $f_i$, $i \in [1\!:\!HT]$; and (ii) $\mu$ is a nonnegative \hbox{trade-off} parameter given as input. The probability $P(X \geq i)$ in~\eqref{symmetric_optimization} can be computed through Proposition~\ref{prop:pmf}. The problem in~\eqref{symmetric_optimization} aims to: (i) maximize the average information rate of MC; and (ii) offer a graceful performance degradation. For $\mu = 0$, the objective function reduces to $R_{\mathrm{MC}}$ in~\eqref{eq:avg_rate_mc}. In this case, due to the constraints in~\eqref{symmetric_optimization}, an optimal solution will select (i.e., assign a nonzero packet fraction) a single erasure code that has the highest average rate. However, this solution does not offer a graceful performance degradation. As $\mu$ increases, an optimal solution allocates nonzero values to a higher number of packet fractions to decrease the \hbox{$\ell_2$-norm} of $f$. This offers a more graceful performance degradation at the cost of achieving a lower average rate. Thus, there is a \hbox{trade-off} between two objectives and there is no unique optimal solution. The parameter $\mu$ can be tuned according to application requirements. 

\subsection{Low-complexity Coding Scheme}\label{subsec:complexity_reduction}
As our scheme combines $HT$ erasure codes, the code complexity increases as $HT$ increases. We propose to reduce the complexity by selecting only $m < HT$ erasure codes. These codes can be selected according to application requirements; here we select them by leveraging our results in Section~\ref{sec:channel_analysis}. If the approximation in~\eqref{eq:prob_approx} holds, the number of received packets at every $T$ time slots is likely to be $jT$ for $j\in[0\!:\!H]$; thus, we can select $m=H$ erasure codes $(HT,jT)$, $j\in[1\!:\!H]$ (or a subset of them to decrease $m$ further). We then combine only the selected codes in our design. The packet fractions of these $m$ erasure codes can be selected by solving~\eqref{symmetric_optimization}. In what follows, we will refer to this heuristic as MC-RC.

\begin{figure*}
     \centering
     \begin{subfigure}[b]{0.271\textwidth}
         \centering
         \includegraphics[width=\textwidth]{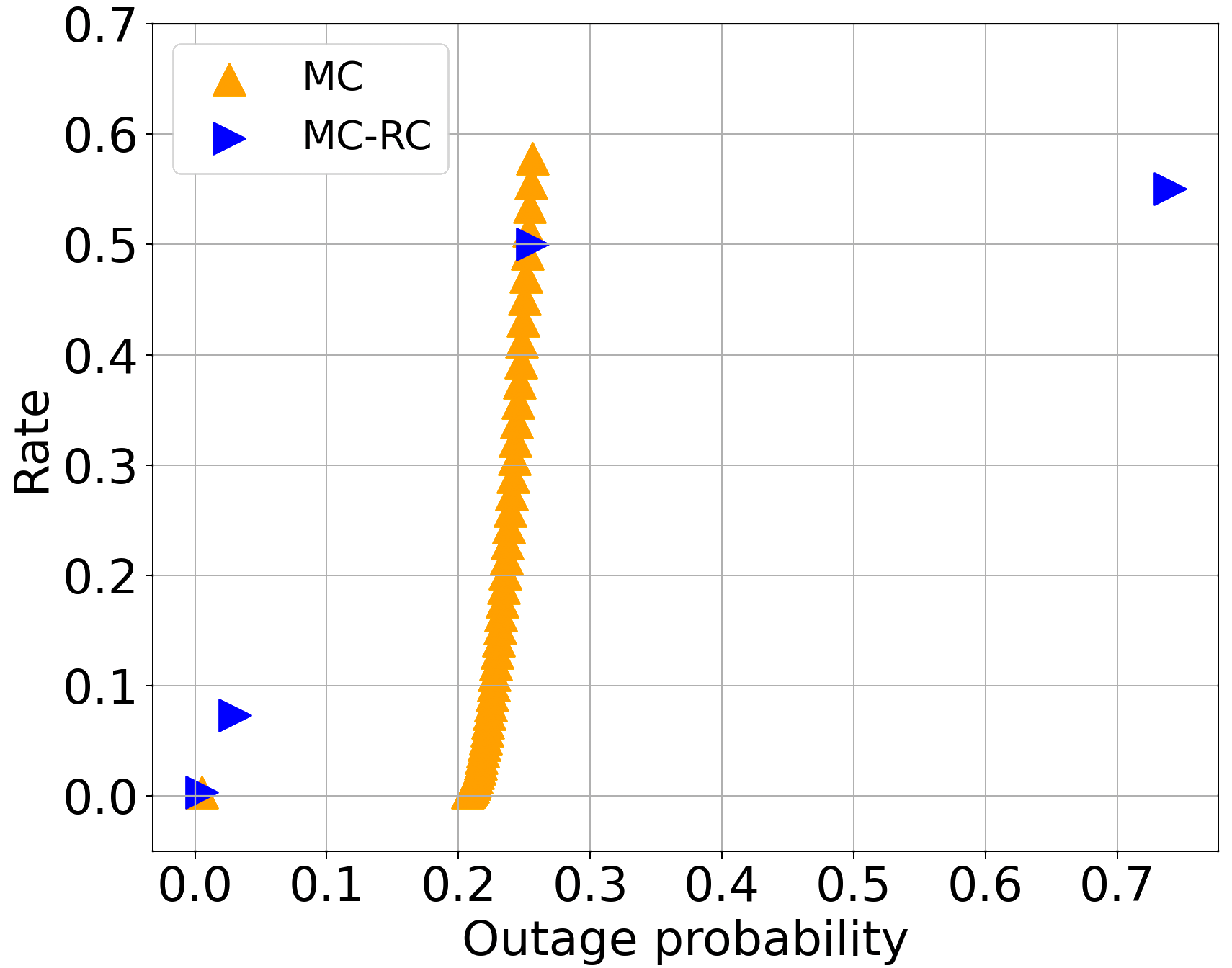}
         \caption{$T=200$ and $L=400$.}
         %Evaluation of the scheduling algorithm.
         \label{fig:rate_vs_outage}
     \end{subfigure}
     \hfill
     \begin{subfigure}[b]{0.266\textwidth}
         \centering
         \includegraphics[width=\textwidth]{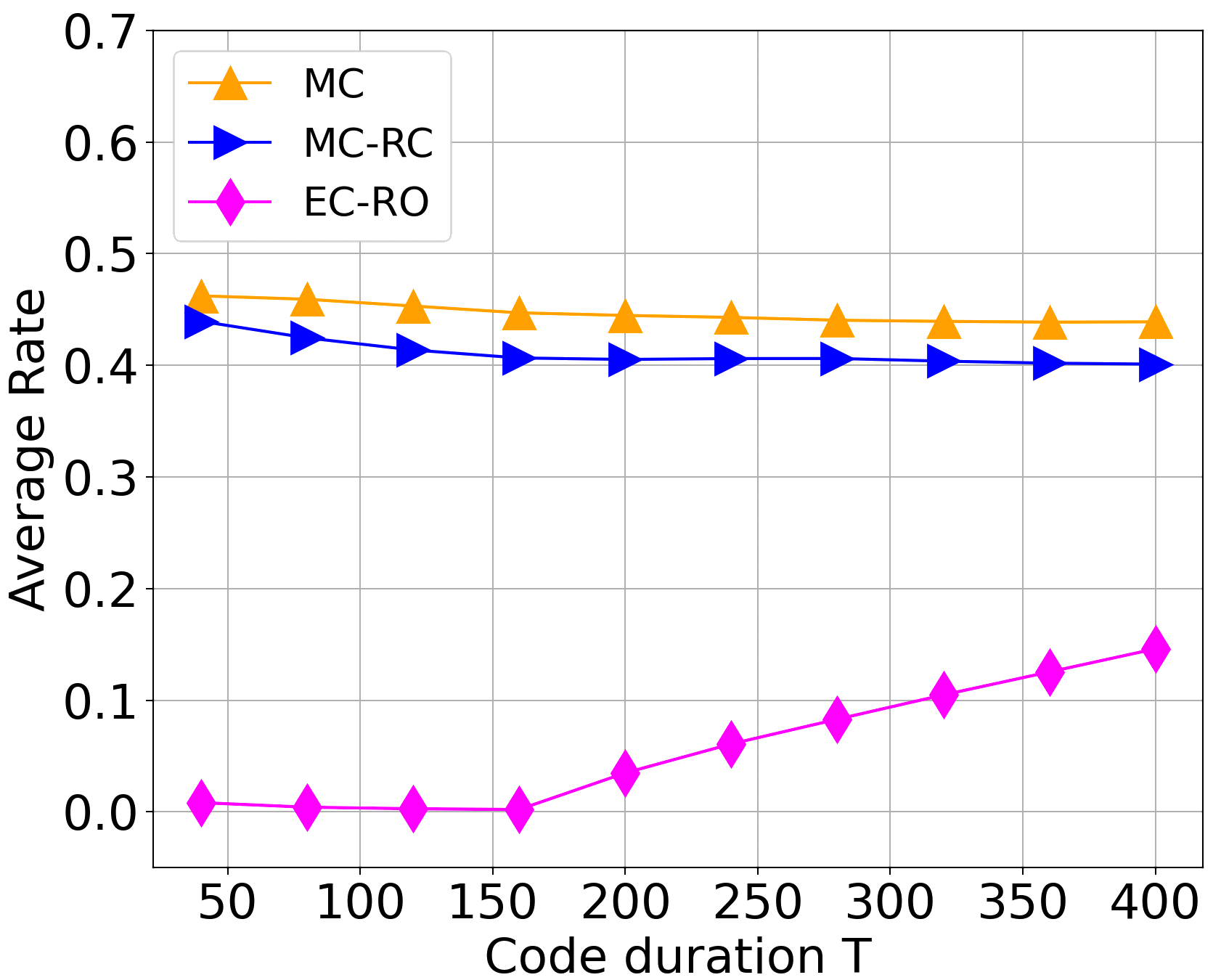}
         \caption{$L=400$.}
         %Evaluation of the path selection algorithm.
         \label{fig:rate_vs_T}
     \end{subfigure}
     \hfill
     \begin{subfigure}[b]{0.268\textwidth}
         \centering
         \includegraphics[width=\textwidth]{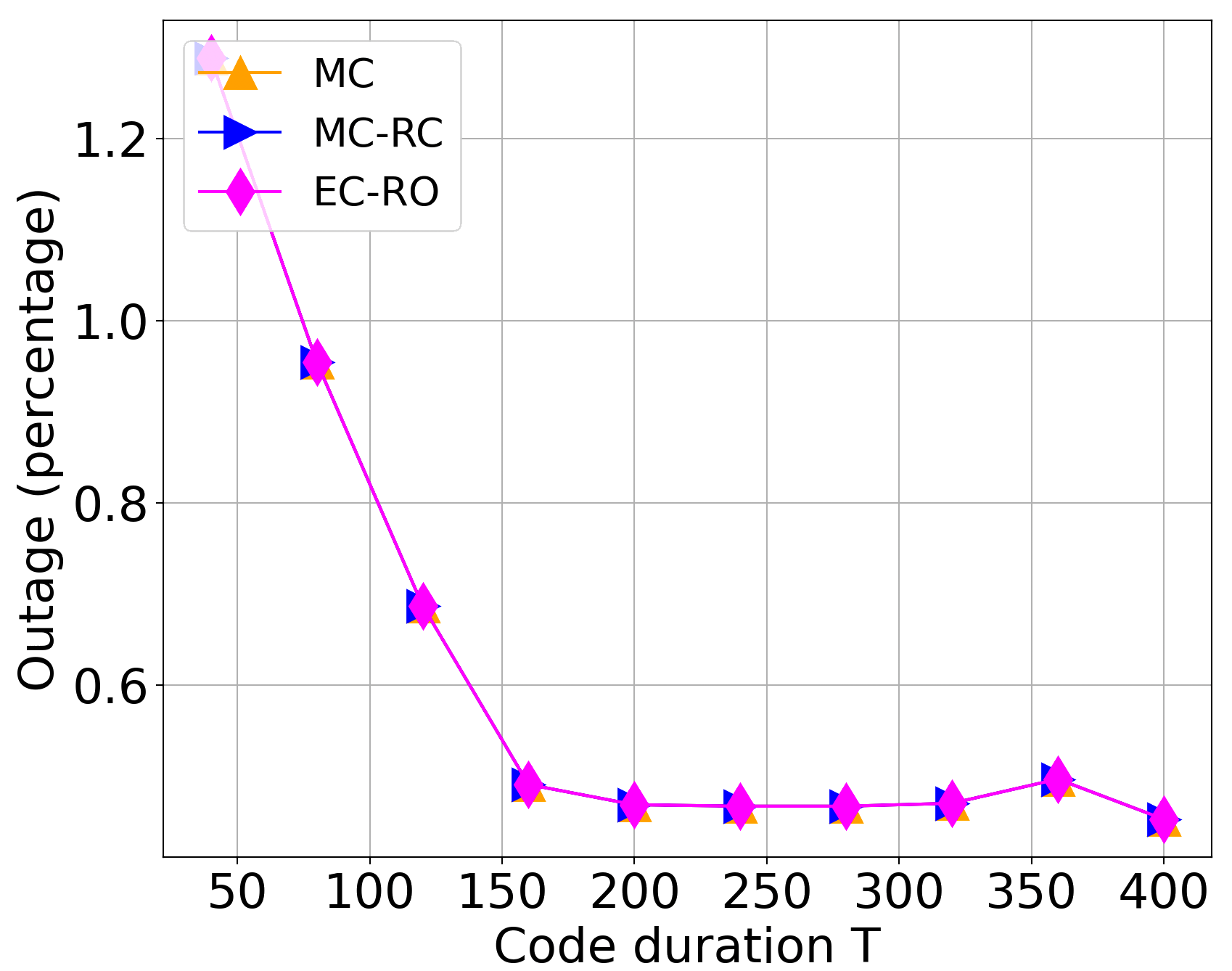}
         \caption{$L=400$.}
         %Evaluation of the path selection algorithm.
         \label{fig:outage_vs_T}
     \end{subfigure}
     \caption{Performance of the coding schemes over the network in Fig.~\ref{fig:example_network1} with $H=3$ \mbox{edge-disjoint} paths.}
        %\label{}
     \vspace{-0.2in}
\end{figure*}

\section{Performance Evaluation}
\label{sec:NumericalEvaluation}
In this section, we assess the performance of our schemes \hbox{MC} and \hbox{MC-RC} with respect to the average information rate, delay, and outage probability. We compare their performance with an alternative scheme, \mbox{\textit{erasure code-reduced outage}} ({\em EC-RO}). \hbox{EC-RO} encodes the source sequences over paths $p_{[1:H]}$ and over $T$ time slots by using a single erasure code. The code is selected such that the outage probability in~\eqref{eq:Pout} is not larger than a given threshold $\gamma$. If there are multiple erasure codes that satisfy this condition, \hbox{EC-RO} selects the code that has the highest information rate among them. The information rate of the selected code is denoted by $R_{\text{EC-RO}}$. If all erasure codes have an outage probability greater than $\gamma$, \hbox{EC-RO} selects the code $(HT,1)$, which has the smallest outage probability. 

We deploy \hbox{MC}, \hbox{MC-RC}, and \hbox{EC-RO} over the network in Fig.~\ref{fig:example_network1}. Our coding schemes can be applied to networks with arbitrary topologies by selecting \mbox{edge-disjoint} paths among all paths. Thus, it can be assumed that the paths in Fig.~\ref{fig:example_network1} are selected from a larger network with an arbitrary topology.
%We compare the performance of our proposed coding schemes with the following alternative scheme\footnote{For all discussed methods, the transmission duration of $HT$ packets is equal to $Tt_d$.}. 
%
%\noindent\textbf{1) Erasure Code (EC).}  
%This method uses a single erasure code, particularly the one with the highest average rate. As discussed in Section~\ref{subsec:mc_optimization}, this is the solution of the optimization problem in~\eqref{symmetric_optimization} for $\mu = 0$. 
%\noindent \textbf{Erasure Code-Reduced Outage (EC-RO).} 
%

We start with the rate and outage probability \mbox{trade-off}. We assume \hbox{$T=200$} and TTI duration $t_d = 250~\mu$s~\cite{Rinaldi21}, thus the transmission delay of $HT=600$ packets is $Tt_d=50$ ms for each scheme. For $L=400$, the blockage duration due to a single blocker\footnote{Overlapping blockage events extend the total blockage duration.} is $Lt_d=100$ ms. The blocker arrival process on each path is a PPP with intensity\footnote{That is, $\alpha_{j} = 7.5 \times 10^{-4}$ blockers per TTI, $j\in[1\!:\!H]$ in~\eqref{eq:eps_alpha}.} $3$ blockers per second. We have source sequences with different priorities and we require that the most important source sequence has to be decoded with a high probability of at least $0.995$. Additionally, we require that it is decoded at least at information rate $R$. We accommodate these requirements by selecting \hbox{$\gamma=0.005$} for \hbox{EC-RO}. \hbox{EC-RO} selects an erasure code $(600,21)$. It achieves rate $R_{\text{EC-RO}}=0.035$ whenever at least $21$ packets are received. We select $R=0.1R_{\text{EC-RO}}$ in this experiment, so that we can support the rate $R$ with $0.005$ outage probability. Similarly, we design \hbox{MC} and \hbox{MC-RC} such that the most important source sequence can be decoded at least at information rate $R$ with \hbox{$P_{\rm{out}}\leq \gamma$}. We do this by ensuring that both \hbox{MC} and \hbox{MC-RC} select \hbox{$f_{21} \geq R/R_{\text{EC-RO}}$} in~\eqref{symmetric_optimization}\footnote{This can be achieved by adding an additional constraint to~\eqref{symmetric_optimization}.}. In Fig.~\ref{fig:rate_vs_outage}, we show the information rate achieved by our schemes versus the outage probability (i.e., the probability that the scheme does not achieve that rate). %For example, the outage probability corresponding to rate $0.50$ of \hbox{MC-RC} is $0.26$; that is, the probability that \hbox{MC-RC} does not achieve rate $0.50$ is $0.26$. 
In Fig.~\ref{fig:rate_vs_outage}, both \hbox{MC} and \hbox{MC-RC} can decode the most important source sequence at rate $R$ with probability $0.995$ (i.e., $0.005$ outage probability). They can decode additional source sequences at higher rates at the cost of having a higher outage probability for them, e.g., \hbox{MC-RC} can decode at least the three most important source sequences at rate $0.50$ with probability $0.74$ (i.e., $0.26$ outage probability). We note that \mbox{MC-RC} combines only $m=4$ erasure codes while MC combines $52$ codes. We also note that \hbox{EC-RO} does not provide different reliability guarantees, and hence it does not exhibit a graceful performance degradation: it either decodes all sequences at rate $R_{\text{EC-RO}}=0.035$, or it fails to decode any of them with probability \hbox{$0.005$}. Thus, a single erasure code only gives a single QoS point, while multilevel codes give a series of points that can suit different QoS requirements of different data streams.

We next evaluate the average information rate and delay \mbox{trade-off}. In Fig.~\ref{fig:rate_vs_T}, we show how the average rate changes as $T$ increases from $40$ to $400$ (i.e., delay increases from $10$~ms to $100$~ms). We find the average rate over the simulated blockage realizations for the network in Fig.~\ref{fig:example_network1} and over $10^8$ time slots. The blocker arrival process on each path is a PPP with intensity\footnote{That is, $\alpha_{j} = 7.5 \times 10^{-4}$ blockers per TTI, $j\in[1\!:\!H]$ in~\eqref{eq:eps_alpha}.} $3$ blockers per second. All schemes encode and transmit source sequences over $p_{[1:H]}$ paths at every $T$ time slots. In Fig.~\ref{fig:rate_vs_T}, we average over the rates achieved at every $T$ time slots\footnote{The achieved information rate depends on the number of received packets over $T$ time slots, on the selected erasure codes, and on the packet fractions.}. For every $T$, \hbox{EC-RO} aims at selecting an erasure code for $\gamma=0.005$. Similarly, both \hbox{MC} and \hbox{MC-RC} are designed for $\gamma=0.005$ and $R=0.1R_{\text{EC-RO}}$. In Fig.~\ref{fig:rate_vs_T}, the average rate of \mbox{EC-RO} decreases until $T=120$ because all erasure codes have outage probability larger than $\gamma$ until $T=120$ and thus, the code $(HT,1)$ is selected by \mbox{EC-RO}. For higher values of $T$, \mbox{EC-RO} selects codes with \hbox{$P_{\rm{out}}\leq \gamma$}. Moreover, outage probabilities of \mbox{low-rate} codes decrease as $T$ increases, which increases their average rates. Since \mbox{EC-RO} selects \mbox{low-rate} codes to satisfy \hbox{$P_{\rm{out}}\leq \gamma$}, its average rate increases. On the contrary, outage probabilities of \mbox{high-rate} codes that are combined by \hbox{MC} and \hbox{MC-RC} increase as $T$ increases. Thus, the average rate of \hbox{MC} and \hbox{MC-RC} decreases.
%\textcolor{blue}{MD: The value of $\varepsilon$ is small and as $T$ increases we send more packets, thus it is possible that low-rate codes don't experience outage. This is because until a blockage event starts, the destination receives enough packets to decode the code. For high-rate codes, this may not hold because even if $T$ and the number of packets increase, it is possible that a blockage event will start and once it starts, we lose the remaining packets due to $L \geq T$. For example, consider rate $1/2$ and $1/40$ codes. If we transmit $HT=40$ packets, we require $20$ and $1$ packets to be received for these codes to be decoded. It is likely that the blockage doesn't start in the first $20$ time slots because $\varepsilon$ is small. Thus, we might decode both codes. However, if we send $400$ packets (and we keep the epsilon the same), we require $200$ and $10$ packets to be received for these codes to be decoded. It is likely that the blockage doesn't start in the first $10$ time slots but it can start before the $200$th time slot which will cause outage for $1/2$ rate code. I couldn't go into details of this because I couldn't create enough space.}

In Fig.~\ref{fig:outage_vs_T}, we show how the percentage of outages changes with $T$ for the code designs in Fig.~\ref{fig:rate_vs_T}. 
%For every $T$, we find the outage percentage as follows. 
Over $10^8$ time slots, at every $T$ time slots we check if outage occurs\footnote{Outage here refers to the case when the most important source sequence cannot be decoded.}. We then plot the percentage of outage events. As shown in Fig.~\ref{fig:outage_vs_T}, all schemes have the same outage percentage as $T$ increases since they use the same erasure code with \mbox{$P_{\rm{out}} \leq \gamma$} (\hbox{MC} and \hbox{MC-RC} use additional codes to improve the rate). Up to $T=120$, all erasure codes have an outage probability larger than $\gamma$. Thus, for these values of $T$, the code $(HT,1)$ is used whose outage probability decreases as $T$ grows. 
%This explains the decrease in the outage percentage. 
As $T$ increases further, there are erasure codes with \hbox{$P_{\rm{out}}\leq \gamma$} and the outage percentage decreases below $0.5\%$.

\section{Conclusions}
\label{sec:Concl}
In this paper, we proposed to deploy multilevel codes both over space and time 
%towards
%we took a first step 
to develop \mbox{low-complexity} proactive transmission mechanisms that offer resilience against link blockages in mmWave networks. 
%In particular, . 
%We formulated an optimization {\color{magenta}problem 
%to select our design parameters 
%such that} we can achieve an attractive \mbox{trade-off} between the average information rate and a graceful performance degradation. 
Our evaluations show that our proposed schemes achieve attractive \mbox{trade-offs} between rate, delay, and outage probability by providing a more graceful performance degradation compared to the alternative scheme, while significantly reducing the complexity.

\bibliographystyle{IEEEtran}
%\balance
\bibliography{IEEEabrv,references}
\appendices
\section{Related Work}\label{appendix:related_work}
There exist several works in the literature that offer resilience against link outages in mmWave networks by taking reactive approaches~\cite{Jeong,Barati,BaratiAsilomar}. In~\cite{MineMilcom}, to achieve resilience to link blockages, the authors explored a \mbox{state-of-the-art} Soft Actor-Critic deep reinforcement learning algorithm, which adapts the information flow through the mmWave network without using knowledge of the link capacities or network topology. However, such reactive mechanisms add the complexity of identification and adaptation, as well as feedback latency. Several works have proposed proactive approaches for resilience by constantly tracking users using side-channel information or external sensors~\cite{Nitsche,Va}. These solutions have limited accuracy, and possibly require sensitive information, such as user location. In~\cite{Mine22,Mine23Arxiv}, the authors leveraged scheduling properties of the mmWave links as well as the blockage asymmetry to achieve the average and the \hbox{worst-case} approximate capacities and proactively offer resilience. They are different from our work where we propose coding schemes to control what information is received, and to accommodate different reliability requirements of different information streams. In~\cite{MineGlobecom}, the authors proposed \mbox{low-complexity} proactive transmission mechanisms for mmWave networks by deploying symmetric multilevel codes over space (i.e., across multiple paths). They proposed an optimization formulation to suitably balance the rate with a graceful performance degradation. The authors then extended this work to scenarios in which the path blockage probabilities are unequal~\cite{MineMILCOM23}. They defined the rate region to operate along with an optimization formulation to select a high-performing set of rates for the source sequences. These works are different from our work because: (i) the operation of their proposed coding schemes relies on the existence of a multipath environment, which might not always be practical; and (ii) these works focus on the \mbox{rate-outage probability} \mbox{trade-off} without considering delay requirements. On the contrary, in this paper: (i) we propose to deploy multilevel codes \textit{both} over space and time, which enables us to deploy the proposed coding schemes even if the network does not support a multipath environment; (ii) our design incorporates time correlation of blockages; and (iii) we consider the \mbox{trade-off} between the rate, delay, and outage probability. 

\section{Proof of Proposition~\ref{prop:pmf}}\label{appendix:proof_pmf_prop}
\begin{figure}
	\centering
    \includegraphics[width=.8\columnwidth]{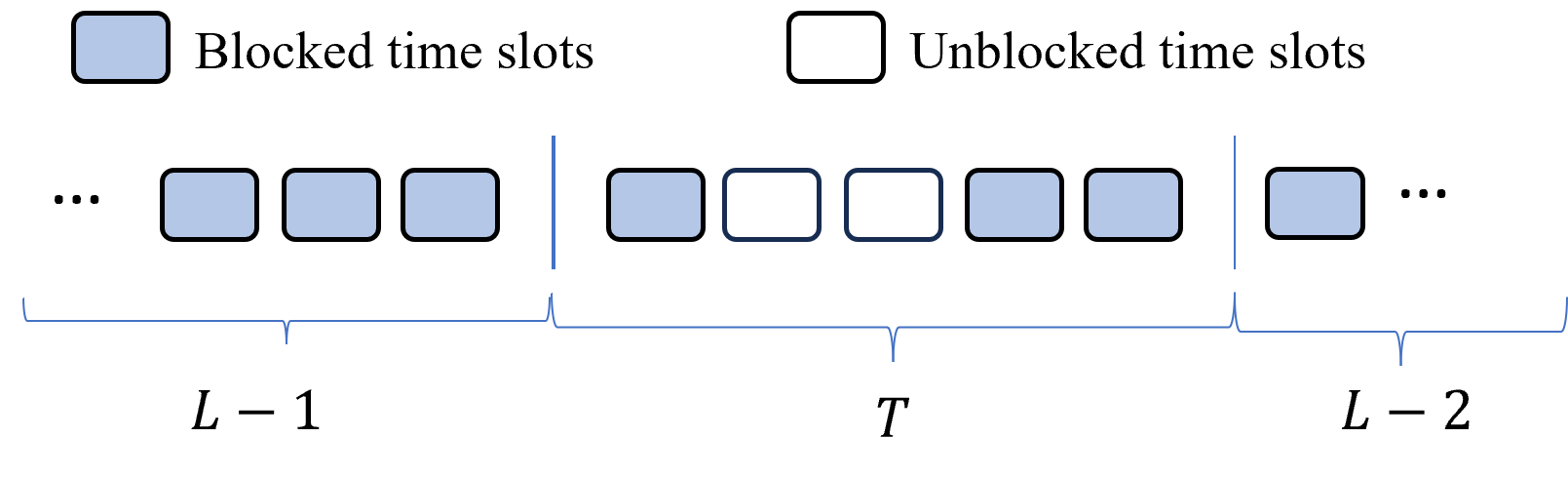}
     \caption{Illustration of time slots for $T=5, r=2$, and $i=1$.}
     \label{fig:prob_cal}
     \vspace{-0.15in}
\end{figure}

%We here derive the PMF $P_{X_r}$, $r\in [1\!:\!H]$ presented in %Proposition~\ref{prop:pmf}. In particular, 
We consider blocks of $T$ time slots as shown in Fig.~\ref{fig:prob_cal}, and we derive the PMF of the number of received packets over these time slots on path $p_j, j \in [1:H]$ (out of the $T$ transmitted packets). 
We note that, once a blockage event starts on path $p_j$, $L$ consecutive time slots are blocked. 
That means, while deriving the probabilities, we need to consider $L-1$ time slots from the previous block as well, as illustrated in Fig.~\ref{fig:prob_cal}.

\noindent {\bf Case~1: $0<r \leq T$.}
%We start with the case $0<k \leq T$. 
The event $\{X_j=r\}$ represents the case where exactly $r$ packets (out of the $T$ transmitted packets) are received, i.e., $r$ out of the $T$ time slots are unblocked. 
Thus, the probability of the event $\{X_j=r\}$ is equal to the sum of the probabilities of the events in which exactly $r$ time slots are unblocked. 
In particular, each of these events can be written as follows. 
In the first $r+i$ time slots (over the $T$ time slots that we consider) for $i\in[0:T-r]$, a new blockage event does not begin. If $i<T-r$, a blockage event begins in the $(r+i+1)$th time slot and the path $p_j$ stays blocked for the remaining $T-(r+i)$ time slots (i.e., the packets transmitted during these time slots are lost) since $L \geq T$. In order to receive exactly $r$ packets we therefore need that: (i) a new blockage event should not start over the last $L-1-i$ time slots in the previous block; and (ii) a blockage event needs to begin in the last $(L-i)$th time slot in the previous block for $i>0$. Thus, the probability of this event is equal to,
\begin{equation}
(1-\varepsilon_j)^{r+i} \varepsilon_j^{\min\{T-r-i,1\}} (1-\varepsilon_j)^{L-1-i} \varepsilon_j^{\min\{i,1\}}.
\end{equation}
Since the probability of the event $\{X_j=r\}$ is equal to the sum of the probabilities of the aforementioned events for \hbox{$i\in[0:T-r]$}, we obtain the probability presented in Proposition~\ref{prop:pmf}.

\noindent
{\bf{Case~2: $r=0$.}}
%We next consider the case $k=0$, i.e., the event $\{X_j=0\}$. 
The probability of the event $\{X_j=0\}$ is equal to the sum of the probabilities of the events in which all the $T$ time slots are blocked. In particular, these events can be due to one of the following events:
\begin{itemize}
\item A new blockage event starts in the first time slot with probability $\varepsilon_j$. Then since $L \geq T$, all $T$ time slots will be blocked independently of what happens in the previous or following time slots.

\item A new blockage event does not start in the first $i$ time slots for $i \in [1:T]$. Then, a new blockage event starts in the $(i+1)$th time slot if $i < T$. In this case, there should be at least one blockage event that started in one of the $L-i$ time slots of the previous block.
This ensures that all $T$ time slots are blocked. Thus, the probability of this event is equal to,
\begin{equation}
(1-\varepsilon_j)^{i} \varepsilon_j^{\min\{T-i,1\}}\left(1-(1-\varepsilon_j)^{L-i}\right).
\end{equation}
\end{itemize}
Since the probability of event $\{X_j=0\}$ is equal to the summation of the probabilities of these events for $i\in[1:T]$, we obtain the probability presented in Proposition~\ref{prop:pmf}. This concludes the proof of Proposition~\ref{prop:pmf}.

\begin{figure*}
     \centering
     \begin{subfigure}[b]{0.26\textwidth}
         \centering         \includegraphics[width=\textwidth]{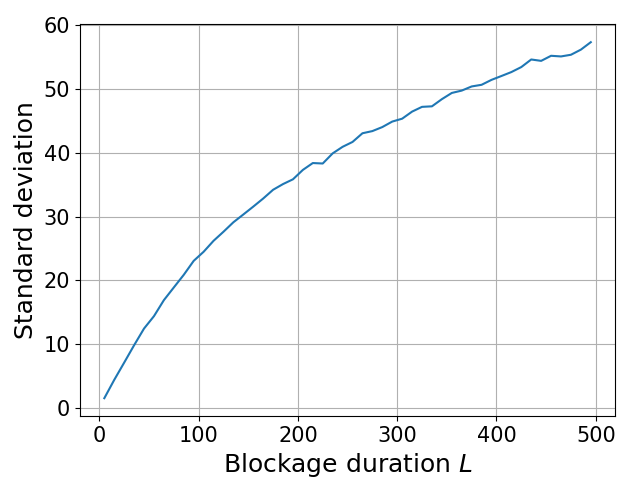}
         \caption{$H=3$}
         %Evaluation of the scheduling algorithm.
         \label{fig:sd}
     \end{subfigure}
     \hfill
     \begin{subfigure}[b]{0.32\textwidth}
         \centering
         \includegraphics[width=\textwidth]{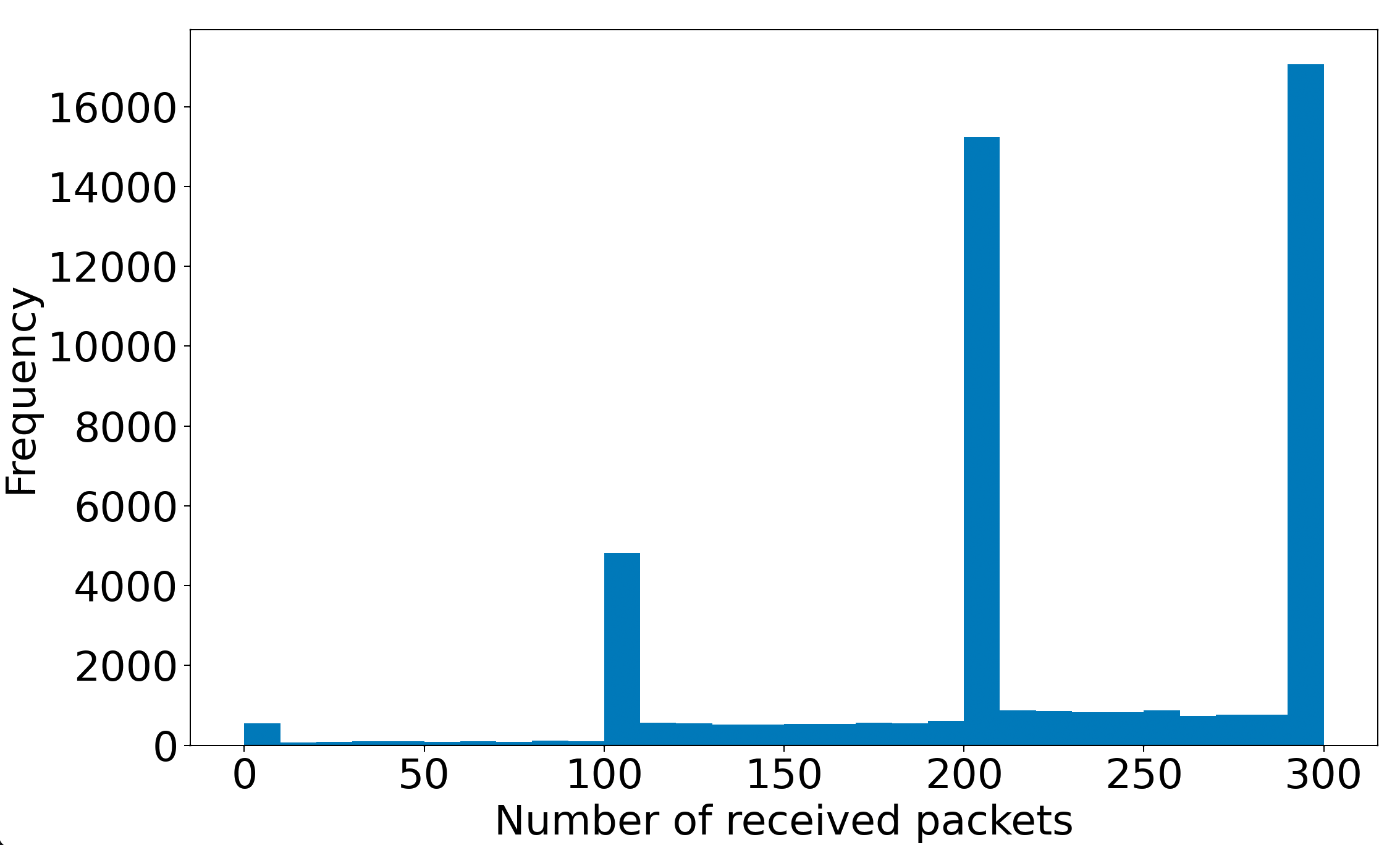}
         \caption{$H = 3, L = 400$.}
         %Evaluation of the path selection algorithm.
         \label{fig:histH3T100L450}
     \end{subfigure}
     \hfill
     \begin{subfigure}[b]{0.32\textwidth}
         \centering
         \includegraphics[width=\textwidth]{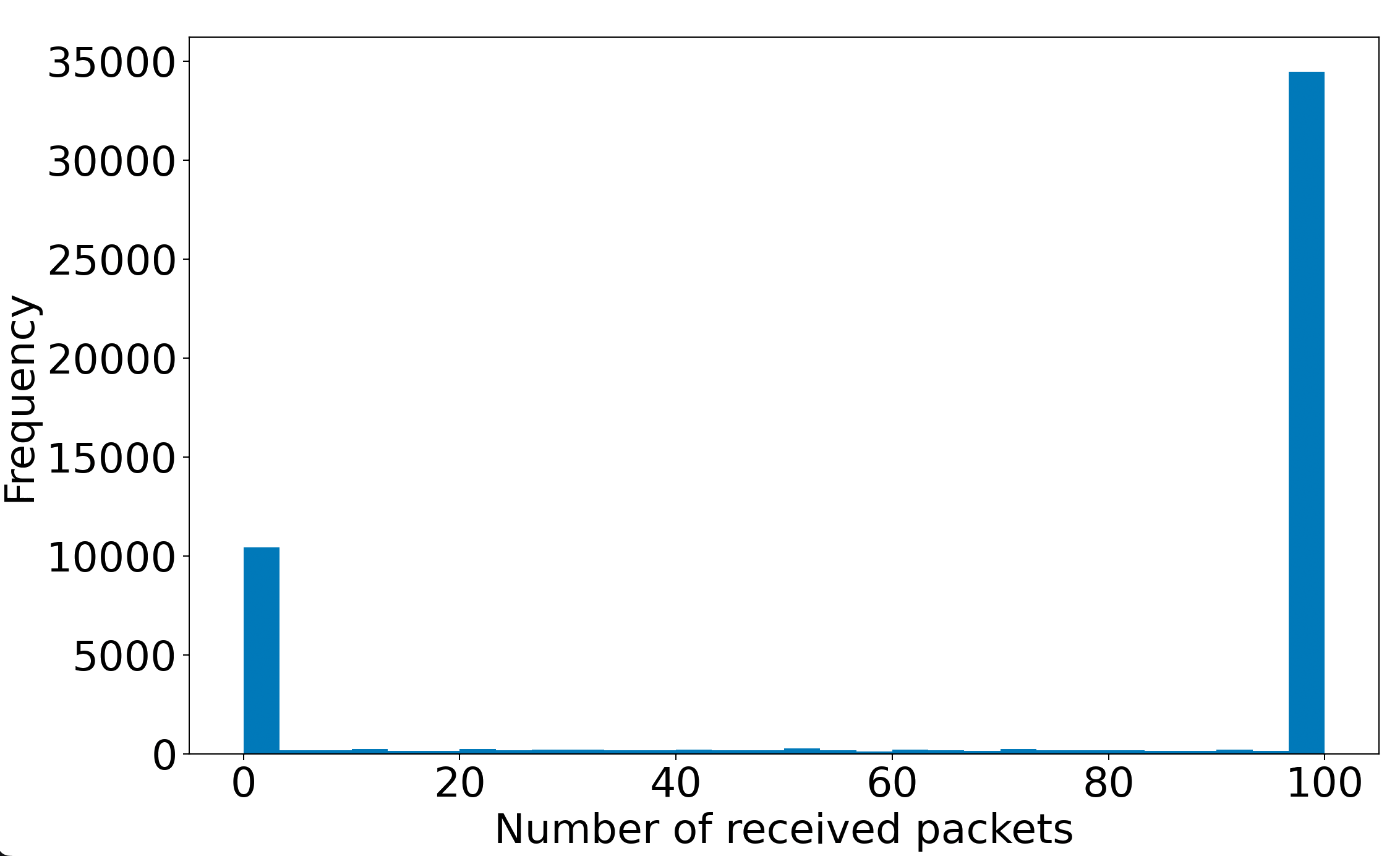}
         \caption{$H =1, L = 400$.}
         %Evaluation of the path selection algorithm.
         \label{fig:histH1T100L300}
     \end{subfigure}
     \caption{Channel illustration for $T=100$, and $A=50,000$.}
        %\label{}
     \vspace{-0.1in}
\end{figure*}
\section{Proof of Proposition~\ref{prop:asymptotic}}
\label{app:ChannelPol}
%We here present the proof of Proposition~\ref{prop:asymptotic}. In particular, 
We start with the probabilities derived in Proposition~\ref{prop:pmf}.
We have that
\begin{align*}
P_{X_j}(0)&=\varepsilon_j+\sum_{i=1}^{T-1} (1-\varepsilon_j)^{i} \varepsilon_j\left(1-(1-\varepsilon_j)^{L-i}\right)
\\ & \quad +(1-\varepsilon_j)^{T}\left(1-(1-\varepsilon_j)^{L-T}\right)
\\& = \varepsilon_j+\varepsilon_j \left(\sum_{i=1}^{T-1} \left((1-\varepsilon_j)^{i}-(1-\varepsilon_j)^{L}\right)\right)
\\& \quad +(1-\varepsilon_j)^{T}-(1-\varepsilon_j)^{L}
\\& = \varepsilon_j+\varepsilon_j\left(\frac{(1-\varepsilon_j)-(1-\varepsilon_j)^T}{\varepsilon_j}-(T-1)(1-\varepsilon_j)^L\right) \\& \quad +(1-\varepsilon_j)^{T}-(1-\varepsilon_j)^{L}
\\& = 1-\varepsilon_j(T-1)(1-\varepsilon_j)^L-(1-\varepsilon_j)^{L}.
\end{align*}
We note that from the above, we readily obtain that
\begin{align*}
\lim_{L \to \infty} P_{X_j}(0) = 1,
\end{align*}
which follows since $\lim_{L \to \infty} (1-\varepsilon_j)^L = 0$.

%If we rearrange the terms, we obtain,
%\begin{align*}
%P_{X_j}(0)= \varepsilon_j+\varepsilon_j \left(\sum_{i=1}^{T-1} \left((1-\varepsilon_j)^{i}-(1-\varepsilon_j)^{L}\right)\right)\\+(1-\varepsilon_j)^{T}-(1-\varepsilon_j)^{L}.
%\end{align*}
%The probability $P(X_j = 0)$ can be equivalently written as,
%\begin{align*}
%P_{X_j}(0)= \varepsilon_j+\varepsilon_j\left(\frac{(1-\varepsilon_j)-(1-\varepsilon_j)^T}{\varepsilon_j}-(T-1)(1-\varepsilon_j)^L\right) \\+(1-\varepsilon_j)^{T}-(1-\varepsilon_j)^{L}
%\end{align*}
%We can simplify the above expression as follows.
%\begin{align*}
%P_{X_j}(0)= 1-\varepsilon_j(T-1)(1-\varepsilon_j)^L-%(1-\varepsilon_j)^{L}.
%\end{align*}
Similarly, from Proposition~\ref{prop:pmf} we have that,
\begin{equation*}
P_{X_j}(T) =  (1-\varepsilon_j)^{T} (1-\varepsilon_j)^{L-1}.
\end{equation*}
Then, we can write $P(X_j = 0)+P(X_j=T)$ as follows,
\begin{align}
P_{X_j}(0)+P_{X_j}(T) 
& = 1-\varepsilon_j(T-1)(1-\varepsilon_j)^L-(1-\varepsilon_j)^{L} \notag
\\& \quad +(1-\varepsilon_j)^{T} (1-\varepsilon_j)^{L-1} \notag
\\& = 1-\varepsilon_j(T-1)(1-\varepsilon_j)^L \notag
\\& \quad -(1-\varepsilon_j)^L\left(1-(1-\varepsilon_j)^{T-1}\right).
\label{eq:sum_probs}
\end{align}
%If we rearrange the terms, we obtain,
%\begin{align}
%\label{eq:sum_probs}
%P_{X_j}(0)+P_{X_j}(T) = 1-\varepsilon_j(T-1)(1-\varepsilon_j)^L\\ \nonumber
%-(1-\varepsilon_j)^L\left(1-(1-\varepsilon_j)^{T-1}\right).
%\end{align}
Now, the approximation in Proposition~\ref{prop:asymptotic} holds if one of the following two conditions holds: (i) \hbox{$\varepsilon_j(T-1) \ll 1$}, or (ii) \hbox{$(1-\varepsilon_j)^L \ll 1$} and \hbox{$\varepsilon_j(T-1)(1-\varepsilon_j)^L \ll 1$}.

\noindent {\bf{Case~1.}} When $\varepsilon_j(T-1) \ll 1$, we can use the Binomial approximation as follows,
\begin{align*}
(1-\varepsilon_j)^{T-1} \approx 1 - (T-1) \varepsilon_j.
\end{align*}
By using the above approximation inside~\eqref{eq:sum_probs}, we obtain
\begin{align*}
P_{X_j}(0)+P_{X_j}(T)  \approx 1-2\varepsilon_j(T-1)(1-\varepsilon_j)^L.
\end{align*}
If the condition $\varepsilon_j(T-1) \ll 1$ in Proposition~\ref{prop:asymptotic} is satisfied, then $\varepsilon_j(T-1)(1-\varepsilon_j)^L \ll 1$ and hence,
\begin{equation*}
P_{X_j}(0)+P_{X_j}(T) \approx 1.
\end{equation*}

\noindent {\bf{Case~2.}} When \hbox{$(1-\varepsilon_j)^L \ll 1$} and \hbox{$\varepsilon_j(T-1)(1-\varepsilon_j)^L \ll 1$}, the second term in~\eqref{eq:sum_probs} is \hbox{$\varepsilon_j(T-1)(1-\varepsilon_j)^L \ll 1$}.
Similarly, since \hbox{$(1-\varepsilon_j)^L \ll 1$}, also the last term in~\eqref{eq:sum_probs} is $(1-\varepsilon_j)^L\left(1-(1-\varepsilon_j)^{T-1}\right) \ll 1$. Hence, 
\begin{equation*}
P_{X_j}(0)+P_{X_j}(T) \approx 1.
\end{equation*}
This concludes the proof of Proposition~\ref{prop:asymptotic}.

\section{Numerical Analysis of Channel}
\label{app:NumericalEvalChannel}
We here perform numerical evaluations for the channel analysis presented in Section~\ref{sec:channel_analysis}.
The specific network configuration under consideration is a \mbox{1-2-1} topology, characterized by a code duration of $T=100$ time slots. We assume that TTI duration $t_d = 250~\mu$s~\cite{Rinaldi21}. The arrival of blockers on every path follows a PPP with an intensity of $3$ blockers per second\footnote{That is, $\alpha_{j} = 7.5 \times 10^{-4}$ blockers per TTI for $j\in[1\!:\!H]$ in~\eqref{eq:eps_alpha}.}. We observe that for this Poisson intensity, we have $\varepsilon_j = 7.5 \times 10^{-4}$. This satisfies the condition $\varepsilon_j(T-1) \ll 1$ in Proposition~\ref{prop:asymptotic}.

First, we consider the network in Fig.~\ref{fig:example_network1} with $H = 3$ \mbox{edge-disjoint} paths. Over this network, we observe the effect of a multipath environment under the considered blockage model. In Fig.~\ref{fig:histH3T100L450}, we present the histogram representing the distribution of the number of received packets for the blockage duration $L=400$ time slots. In order to plot this histogram, we simulate blockage realizations for the network in Fig.~\ref{fig:example_network1} and over $A=50,000$ time slots in total. We find the number of received packets at every $T$ time slots, and we plot their distribution in~Fig.~\ref{fig:histH3T100L450}. We note that the total number of transmitted packets at every $T$ time slots is $HT = 300$.

As discussed in Section~\ref{sec:channel_analysis}, when one of the conditions in Proposition~\ref{prop:asymptotic} is satisfied and there are $H$ paths, the number of received packets is likely to take values $jT$ for $j\in[0:H]$. This concentration is observable in Fig.~\ref{fig:histH3T100L450}, where a considerable proportion of mass is evident at intervals of $T$, such as $0,T,2T,3T$. Consequently, this concentration contributes to an increased variance in the channel behavior, which motivates the use of multilevel coding in a multipath scenario. In Fig.~\ref{fig:sd}, we show how the standard deviation of the number of received packets from the mean changes as the blockage duration (in time slots) $L$ increases. As $L$ increases, we observe this concentration around specific values $jT$ for $j\in[0:H]$, as mentioned above. And due to that, standard deviation also increases. 

To numerically verify Proposition~\ref{prop:asymptotic}, we also consider the case for $H=1$. In Fig.~\ref{fig:histH1T100L300}, we observe consistent results with our analysis in Proposition~\ref{prop:asymptotic}. Specifically, the distribution is primarily concentrated around $0$ and $T$, allowing us to numerically confirm that \hbox{$P_{X_1}(0)+P_{X_1}(T) \approx 1$}.

\end{document}